# Full-Field Metasurface Characterization with Polarization Sensitive Coherent Modulation Imaging

Xinjie Sun[1], Xin Liu[2], Zixin Cai[1], Yanghui Li[3, *], Xu Liu[1, 4, *], and Xiang Hao[1, 4, *]

[1] *State Key Laboratory of Extreme Photonics and Instrumentation, College of Optical Science and Engineering, Zhejiang University, Hangzhou 310027, China*

[2] *Department of Electrical and Electronic Engineering, The University of Hong Kong, Hong Kong SAR, China*

[3] *College of Optical and Electronic Engineering, China Jiliang University, Hangzhou 310018, China*

[4] *ZJU-Hangzhou Global Scientific and Technological Innovation Center, Hangzhou 311200, China.*

*\* haox@zju.edu.cn; liuxu@zju.edu.cn; lyh@cjlu.edu.cn*

**Abstract:** Characterizing the intensity, phase, and polarization of engineered light is fundamental to understanding and applying metasurfaces. However, existing characterization frameworks are hindered by several limitations, most notably their inability to account for the polarization of the field. Here, we report polarization sensitive coherent modulation imaging (PS-CMI), a light-weight but robust, high-resolution platform for the full-field characterization of metasurface-modulated light. By supplementing the orthogonal *x*- and *y*- complex amplitude components with an additional 45°-component, this approach calculates the retardance between two orthogonal polarization components while eliminating phase offsets, thereby enabling the subsequent recovery of the complete polarization state. We demonstrate the versatility of our method by characterizing light fields produced by a United States Air Force (USAF) target, two kinds of complex polarization field, and a metalens. This compact solution addresses a critical gap in metasurface metrology and is broadly applicable to other fields requiring the mapping of complex, polarized light distributions.

## 1. Introduction

Metasurfaces, the planar optical architectures consisting of subwavelength optical antenna arrays (*1*), have emerged as a powerful paradigm. They can simultaneously control the fundamental degree of freedom in light, including phase, amplitude, and polarization. By consolidating complex wavefront manipulation into a single, ultra-thin layer, metasurface devices offer a compelling alternative to conventional bulk optics, such as lenses, waveplates, and polarizers (*2–4*), particularly for weight- and compactness-sensitive applications in mobile telephony and autonomous navigation. However, fully unlocking the potential of these devices necessitates a simultaneous, full-field, and high-resolution recovery of all modulated properties in the emergent light field, which remains elusive.

Metasurface-modulated light field characterization has long relied on interferometry as the standard benchmark (*5*, *6*). However, its implementation is frequently hampered by a reliance on complex, dual-path optical architectures. These configurations are notoriously sensitive to environmental perturbations such as mechanical vibrations and thermal drifts, which degrade measurement stability. In contrast, coherent diffraction imaging (CDI) methods, such as coherent modulation imaging (CMI), Fourier ptychographic microscopy (FPM), and multi-distance CDI, have emerged as robust, common-path alternatives (*7*, *8*). However, current CDI frameworks for metasurface characterization are restricted to scalar wave-field recovery. They map only amplitude and phase and fail to resolve the polarization information.

Beyond the specific domain of metasurface characterization, several CDI-derived modalities have sought to bridge this gap by recovering both the complex amplitude and the polarization state of optical beams. One strategy integrates polarization-sensitive detectors into FPM (*9, 10*), though often at the expense of spatial resolution. Alternatively, structured illumination has been employed to reconstruct the full vectorial distribution of tightly focused fields in three-dimensions (3Ds) (*11*). While such illumination encoding promotes the resolution, it compromises the ability to characterize samples sensitive to incident polarization and angle, such as metasurfaces. In contrast, measurement-encoding schemes—for example, combining multi-distance CDI with time-sequential polarization measurement (*12*)—provide a more robust approach for characterizing these samples. Nevertheless, these approaches remain restricted to the paraxial regime, inherently limiting the achievable resolution for characterizing metasurfaces with sub-wavelength structures.

To overcome these challenges, we present polarization sensitive coherent modulation imaging (PS-CMI), a framework for the high-resolution characterization of both wavefront and polarization profiles emergent from metasurfaces. Our new design provides two major improvements over the original works (*7–9, 11–14*):

- Comprehensive characterization and nondestructive diagnosis for metasurfaces: We demonstrate that the phase offset, an intrinsic ambiguity in CDI methods, is not globally consistent and must be addressed through regional analysis. Based on this insight, we develop a phase-offset removal algorithm that accurately retrieves the polarization map. While maintaining consistency with prior works on phase modulation (*7, 8, 14*), our method incorporates reconstructed maps to characterize the metasurface's impact on polarization. Furthermore, reconstructed polarization maps enable nondestructive diagnosis of fabrication errors in metasurfaces, which provide guidance for subsequent design optimization and fabrication refinement.
- Broad Sample Compatibility: By employing measurement encoding rather than illumination encoding, our PS-CMI enables the characterization of angle- and polarization-sensitive samples, a category epitomized by metasurfaces. This framework is readily integrable with conventional microscopy. Rather than compromising the performance, the addition of PS-CMI enhances the resolution of the base system. This renders our method uniquely suited for high-resolution applications and samples that induce large-angle beam modulations.

## 2. Principles

The full field is reconstructed using CMI-based phase retrieval and time-division polarization modulation. Theoretically, the vectorial light field can be reconstructed using Jones vector once the complex amplitudes for the *x*- and *y*-polarization components are established. Due to its ill-posed nature, phase retrieval suffers from an inherent ambiguity manifesting as a phase offset within the CMI framework (*15*). This ambiguity leads to incorrect retardance values between *x*- and *y*- polarization components, and the Jones vector of the reconstructed field ($U'$) containing phase offset can be expressed as:

$$U' = \begin{pmatrix} U_x' \\ U_y' \end{pmatrix} = \begin{pmatrix} u_{x0} e^{i\theta_x} e^{i\theta_\leftrightarrow} \\ u_{y0} e^{i\theta_y} e^{i\theta_\updownarrow} \end{pmatrix} \triangleq \begin{pmatrix} u_{x0} e^{i\theta_x} e^{i\Delta\theta} \\ u_{x0} e^{i\theta_y} \end{pmatrix} \cdot e^{i\theta_\updownarrow} \quad (1)$$

where $\theta_\leftrightarrow$ and $\theta_\updownarrow$ are the phase offsets in *x*- and *y*-directions, respectively. and $\Delta\theta = \theta_\leftrightarrow - \theta_\updownarrow$. To enable phase offset removal, an additional component $U_{45}'$ is introduced to derive a relationship among

$U'_x$, $U'_y$, $U'_{45}$ and $\Delta\theta$ (12). This allows $\Delta\theta$ to be determined via a one-dimensional search (detailed in part A of the Supplementary Materials). Distinguished from the previous methods that treat the phase offset as a global constant, we calculate the phase offset as a partitioned local constant to handle complex polarization fields with discontinuous phases. Further details regarding this method are expounded in the 'Experiments and Results' section.

Fig. 1 illustrates our PS-CMI method for the full-field metasurface characterization. The setup begins with a plane wave illuminating the metasurface (or other light-field modulating elements). The transmitted field then propagates through a rotatable half-wave plate (HWP) and a fixed linear polarizer (LP). By sequentially setting the HWP rotation to 0°, 22.5°, and 45°, we utilize its combination with the LP to selectively detect the 0° (*x*)-, 45°-, and 90° (*y*)-polarization components of the transmitted field, respectively. Next, the light field is modulated by a series of random phase patterns, and iterative phase-retrieval algorithms are employed to independently reconstruct the complex amplitudes of these polarization components. To address the phase ambiguity inherent in independent reconstructions, a phase offset removal algorithm is applied to retrieve the accurate retardance between the *x*- and *y*-components, ultimately enabling the full-field reconstruction of the metasurface.

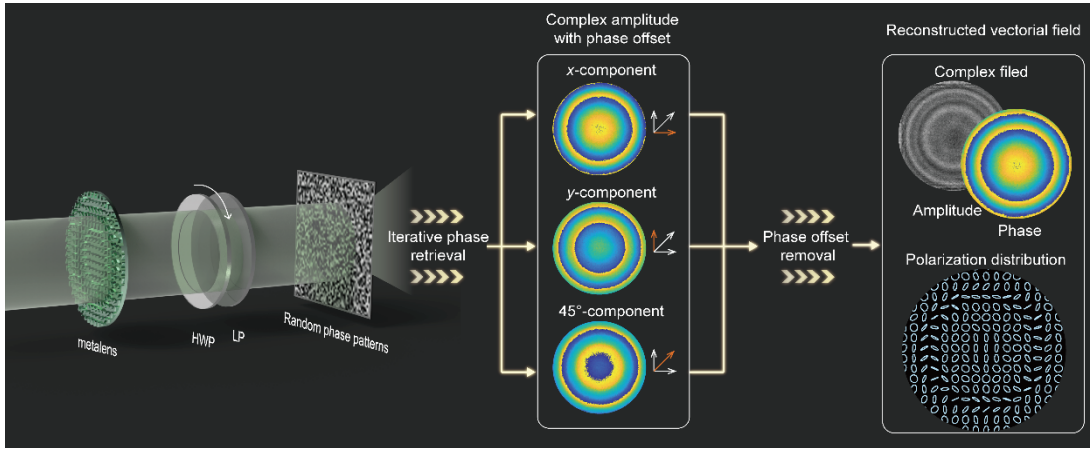

**Fig. 1**. **Schematic of polarization sensitive coherent modulation imaging (PS-CMI) for field reconstruction.** To reconstruct the light field modulated by the metalens, we initially isolate the field into its 0° (*x*), 90° (*y*), and 45° polarization components using a setup comprising a rotatable half-wave plate (HWP) and a fixed linear polarizer (LP). Next, a series of pre-defined random phase patterns are imposed onto each component field, and the resulting diffracted intensity distributions are recorded. An iterative phase-retrieval algorithm is then employed to reconstruct the complex amplitude of each component field. Due to the inherent ambiguities in iterative phase-retrieval algorithms, the reconstructed phase carries a phase offset. We introduce the 45°-component within the phase offset removal algorithm to calculate the phase offset difference ($\Delta\theta$) between the *x*- and *y*-components, thereby yielding the correct phase difference between the two orthogonal directions. The final light field emergent from the metalens is then obtained by the polarization synthesis of the *x*- and *y*-component fields, corrected by the determined phase offset. Abbreviations: HWP, half waveplate; LP, linear polarizer.

## 3. Experiments and results

We integrated the described setup into a microscope to achieve a high spatial resolution (Fig. 2A). The sample plane is optically conjugated with a spatial light modulator (SLM), which displays a series of random phase patterns. The resulting modulated light field is reflected by a beam splitter (BS) and recorded by a complementary metal-oxide-semiconductor (CMOS) sensor. Full details of the experimental setup, SLM–CMOS configuration, and random phase pattern generation are provided in the **Methods** section.

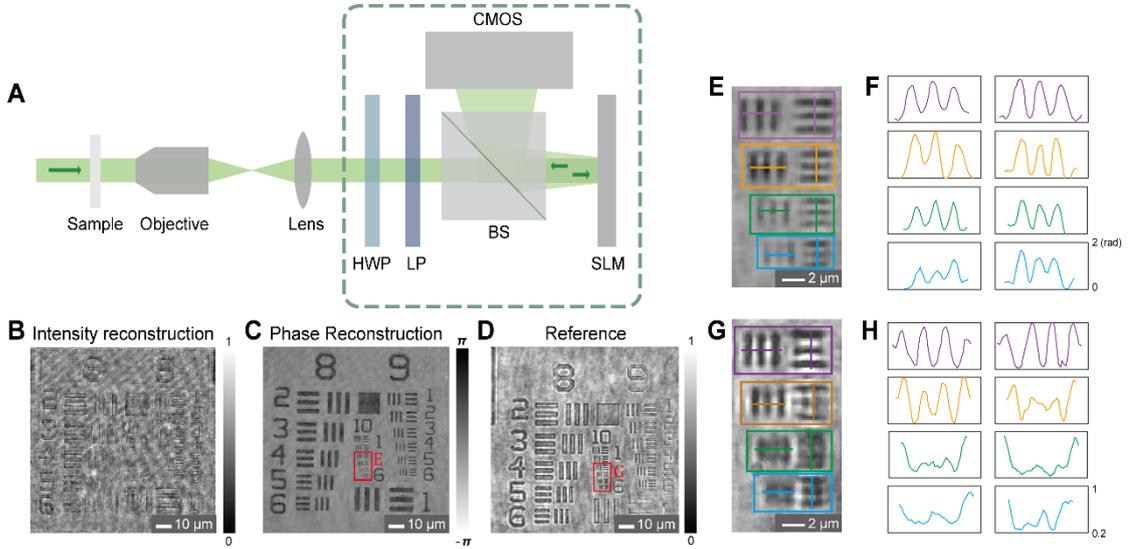

**Fig. 2. System and performance of our method. A,** simplified schematic of the optical system. The dashed box indicates the core setup detailed in Fig. 1. **B,** intensity and **C,** corresponding phase reconstruction of a United States Air Force (USAF) target using our method. **D**, reference image of the same USAF target captured with a bright-field microscope. The red boxes in **C** and **D** indicate the region of interest. **E,** magnified views of the phase reconstruction (from the red box in **C**) and **G,** corresponding magnified view of the reference (from the red box in **D**). **F, H,** line profiles extracted from the colored boxes in **E** and **G**, respectively. The *y*-axis ranges are consistent with the labels in the bottom-right panels; the width of each plot is 2.33 μm. Abbreviations: BS, beam splitter; CMOS, complementary metal-oxide-semiconductor.

To assess the performance of our system, especially its resolution and phase reconstruction fidelity, we characterized a phase-type United States Air Force (USAF) target. This sample features a step height of 250 nm and a refractive index of 1.52. Utilizing a 50× objective, the reconstructed intensity and phase distributions were obtained (Fig. 2B–C). As expected for a phase object, the phase reconstruction exhibits significantly higher signal-to-noise ratio (SNR) than the corresponding intensity one (28.18 vs. 7.55). For validation, a reference intensity image of the same field of view was acquired using a wide-field microscope equipped with the identical objective (Fig. 2D), providing a benchmark for our reconstruction results.

Magnified views of the four-line pairs within the red boxes in Fig. 2C and Fig. 2D are shown in Fig. 2E (phase reconstruction) and Fig. 2G (reference). The corresponding phase and amplitude line profiles along the colored lines in Fig. 2E and Fig. 2G are displayed in Fig. 2F and Fig. 2H, respectively. According to the Rayleigh criterion, the fringes are considered resolvable if the valley intensity is less than 80% of the peak intensity. Thus, our method resolves the target up to Group 10, Element 6, corresponding to a line width of 548.1 nm (Fig. 2E–F). In contrast, the finest resolvable feature for the reference corresponds to Group 10, Element 4, equivalent to a line width of 690.6 nm (Fig. 2G–H). Although measurement encoding ties the system resolution to the diffraction limit of the microscope, these results demonstrate that integrating our PS-CMI method not only preserves but enhances the resolution performance of the original microscope system. This improvement is attributed to the enhanced SNR afforded using multiple phase-pattern encodings.

To further demonstrate the system capability in polarization reconstruction, we characterized three distinct polarized light fields. Radially polarized light (Fig. 3A–C) and azimuthally polarized light (Fig. 3D–F) were generated using a first-order vortex retarder. To generate a more complex polarization field with intricate ellipticity, a QWP (fast axis at 0°) was placed after the radially polarized light (Fig. 3G-I). Fig. 3A, 3D, and 3G illustrate the reconstructed complex amplitudes for the x- and y-components of the three polarized light fields after removing the phase offset. In these images, the brightness corresponds to the amplitude, while the colormap represents the phase.

During polarization reconstruction, we observed that the phase offset should not be treated as globally consistent but should instead be solved regionally. This is primarily because the existing phase-offset algorithm(*12*) has smoothness assumption, whereas the iterative phase-retrieval process inherently introduces phase wrapping. To avoid errors caused by these phase discontinuities, the phase offset must be calculated in regions. For computational efficiency, we use a 128 × 128 regional division as the default for phase offset computation. If higher polarization resolution is needed, this partitioning can be refined to a pixel-wise scale. The detailed discussion of sub-region division is available in Fig. S1 discussed in part B of the Supplementary Materials. Notably, when the regions are partitioned finely enough that the polarization state within each area can be considered uniform, the conventional phase offset algorithm encounters chirality ambiguity. Consequently, while the local phase offset algorithm can resolve the orientation and absolute ellipticity of polarization fields with phase discontinuities, it remains unable to determine the handedness.

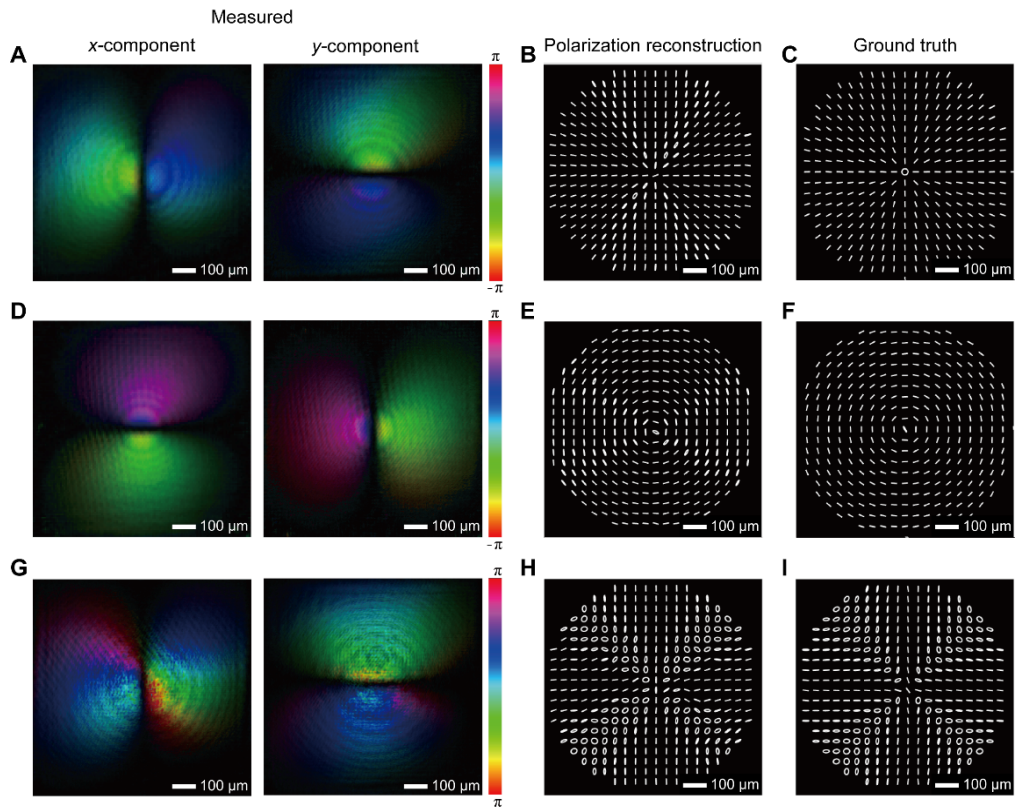

Fig. 3. Polarization reconstruction of cylindrically polarized beams. **A–C**, complex amplitude reconstruction, polarization reconstruction, and the ground truth (obtained from a Stokes camera) of a radially polarized field. **D–F**, complex amplitude reconstruction, polarization reconstruction, and the ground truth of an azimuthally polarized field. **G-H**, complex amplitude reconstruction, polarization reconstruction,

and the ground truth of the field generated by a QWP (fast axis at 0°) placed after the radially polarized light.

The brightness in **A**, **D** and **G** corresponds to amplitude, while the colormap represents the phase.

Following the sub-region division, Fig. 3B, 3E and 3G present the reconstructed polarization maps for three fields, respectively. To validate these results, ground truth distributions were acquired using a Stokes camera (Fig. 3C, 3F and 3I). The reconstructions exhibit high fidelity to the ground truth, capturing both polarization orientation and ellipticity. To quantify the performances, we define a polarization reconstruction accuracy metric based on the mean power overlap. This metric represents the average pixel-wise correlation between the measured values and the ground-truth values (detailed in part C of the Supplementary Materials). Using this definition, the polarization reconstruction accuracies for these three polarization fields reach 94.8%, 95.2% and 93.3%, respectively. However, if a global phase offset is used instead of local phase offsets, the recovered polarization state becomes erroneous (as shown in Fig. S2 of the Supplementary Materials), thereby degrading the reconstruction accuracy to 74% (for the radially polarization field). These results unequivocally demonstrate system capability to accurately reconstruct complex polarization fields.

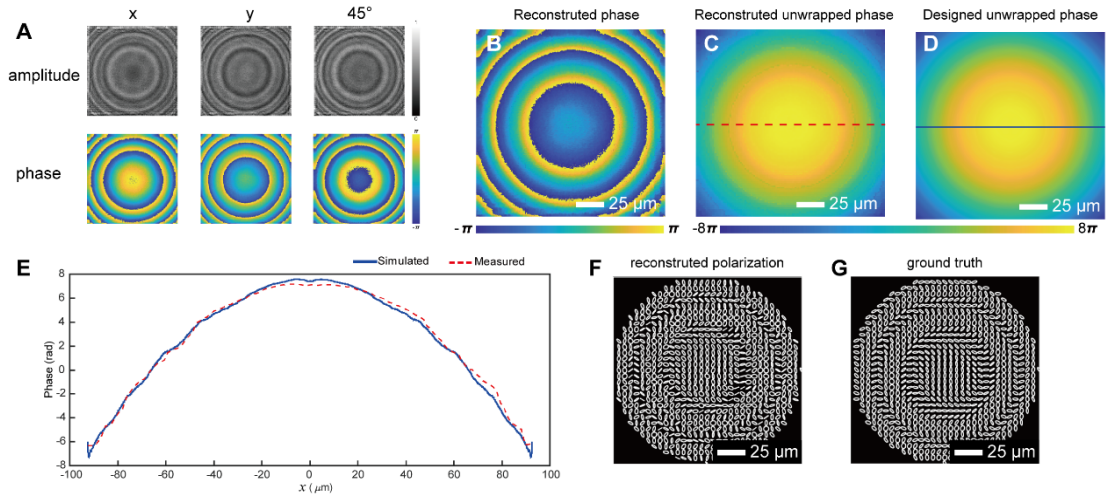

**Fig. 4. Characterization of focusing metalens using our method. A,** phase reconstruction of the *x*-, *y*-, and 45°-components in the metalens's transmitted field. **B,** reconstructed phase of the left-handed component after phase offset removal. **C,** unwrapped phase derived from **B**. **D,** unwrapped phase distribution of the left-handed component in design. **E,** phase profiles at positions denoted by the dashed and solid lines in **C** and **D**, respectively. **F,** reconstructed polarization map obtained with our method. **G,** ground truth measured by a stokes camera.

Characterizing the light fields generated by metasurfaces is inherently a challenge, as these devices simultaneously modulate the complex amplitude and the polarization state. To validate our methodology, we fabricated a geometric-phase metalens designed for $\lambda$ = 532 nm (NA = 0.1; focal length = 3.5 mm). The design and fabrication of the metalens are detailed in the Methods section. This metalens provides a robust test case by combining focusing functionality with a complete handedness reversal of the incident circular polarization.

We illuminated this metalens with a 532 nm, right-handed circularly polarized, collimated continuous-wave laser beam. The amplitude and phase of the transmitted field across three polarization directions were computationally reconstructed (Fig. 4A). Similarly, we also calculated the phase of the crucial left-handed component from the *x*- and *y*-components (Fig. 4B), and the phase was subsequently unwrapped (Fig. 4C) for direct comparison with the theoretical design (Fig. 4D).

Profiles taken from the middle of the reconstructed and designed phase patterns (Fig. 4E) show a remarkable correspondence, yielding a phase reconstruction accuracy of 93.6%. Here, the phase reconstruction accuracy is defined based on normalized root-mean-square-error (*11*). The reconstructed phase shows excellent agreement with the theoretical results, demonstrating that the orientations of the fabricated nanopillars are consistent with the design. The phase of the left-handed component is derived from the *x*- and *y*-components following the removal of the local phase offsets. Consequently, its accuracy is subject to errors from both the iterative phase-retrieval and phase offset removal algorithms. To quantify the intrinsic accuracy of the iterative phase-retrieval algorithm, we independently assessed the phase profile of the *x*-component, yielding an accuracy of 94.5% (see Fig. S3 in Supplementary Materials). This result confirms that the phase offset removal algorithm introduces neglectable error, demonstrating that the final accuracy is dominated by the iterative phase-retrieval.

The light field emergent from the metalens exhibits a dramatically higher spatial complexity than the cylindrical beams discussed previously. Our high-fidelity polarization reconstruction provides deeper insights into device performance. Fig. 4F presents the polarization distribution recovered via our PS-CMI, which demonstrates a reconstruction accuracy of 93.4% relative to the Stokes-camera benchmark (Fig. 4G). This performance is commensurate with the phase accuracy reported in Fig. 4E. It validates the ability of our system to maintain accuracy even across highly heterogeneous fields.

While existing characterization methods for metalenses are typically restricted to scalar metrics, they cannot reveal underlying fabrication errors. In contrast, PS-CMI provides a more nuanced understanding of the polarization response. A comparative analysis between the simulated polarization profiles (as shown in Fig. S4A-B of the Supplementary Materials), the PS-CMI reconstructions, and the Stokes-camera ground truth reveals polarization accuracies of only 83.12% and 83.16%, respectively. These results indicate that while the local orientation is precisely preserved, fabrication-induced morphological variations in the nanopillar geometry (as shown in Fig. S4C of the Supplementary Materials) have compromised the polarization response of the metalens. Consequently, these structural departures from the ideal design significantly attenuate the polarization conversion ratio, which drops from a theoretical value of 4,500:1 to an experimentally realized 513:1. Conventional identification of such defects relies on Scanning Electron Microscopy (SEM), which is inherently destructive. Hence, PS-CMI provides a non-destructive alternative for characterizing metasurface fabrication errors, offering critical guidance for manufacturing optimization and ultimately accelerating the advancement of metasurface technologies.

## 4. Discussion

As illustrated in Fig. 2, our PS-CMI improves the achieved resolution from 690.6 nm to 548.1 nm. This ~20% enhancement originates from the superior SNR facilitated by the multi-frame acquisition inherent to CMI. By employing measurement encoding instead of illumination encoding, we guarantee compatibility with angle- and polarization-sensitive samples, though the system resolution remains fundamentally constrained by the diffraction limit of the objective lens.

Our platform achieves >93% reconstruction accuracy for both phase and polarization reconstruction, with further accuracy gains contingent upon SNR optimization. While elevated brightness in CMI inevitably saturates low-frequency signals, it effectively enhances the acquisition of weak high-frequency signals, albeit at the cost of requiring more phase patterns or iterations (*16*). Consequently, the integration of high-dynamic-range (HDR) detectors would be beneficial, allowing for the reliable capture of the full spectral distribution while mitigating signal saturation.

The results presented here were all obtained in fixed samples. The temporal efficiency of our method is primarily dictated by the number of random phases required. For the intricate fields produced by the metalens, we employed 60 pre-defined random phase patterns to ensure maximum fidelity (as shown in Fig. S5A of the Supplementary Materials). Since this quantity is commensurate with the refresh rate of the SLM (60 fps), the characterization of rapid, dynamic light fields remains a technical challenge. Notably, we observe that the required number of phase patterns scales with the topological complexity of the field. Simpler light distributions can still be characterized using a reduced phase number without compromising reconstruction fidelity. For example, for the azimuthally polarized beam shown in Fig. 3B, 30 patterns are sufficient to achieve 94% accuracy (as shown in Fig. S5B of the Supplementary Materials). For non-polarimetric USAF sample (*e.g.* Fig. 2), the number of patterns can be further reduced to 8 without degrading resolution.

In summary, our PS-CMI method provides a rigorous and versatile framework for the simultaneous, high-resolution recovery of the amplitude, phase, and polarization of complex light fields. A comprehensive benchmarking against other CDI modalities (Table S1 of the Supplementary Materials) highlights the unique capacity of our method to resolve multiple optical parameters in angle- and polarization-sensitive nanophotonic devices such as metasurfaces.

By providing nondestructive access to the fundamental amplitude, phase and polarization properties of the light field, our method is poised to impact diverse applications beyond metasurface metrology and represents an important step towards high-performance computational optics.

## Methods
### Experimental Configuration
The optical setup is schematically detailed in Fig. 2A. Light emergent from the sample is collected by an objective and then collimated via a tube lens (#36-401, Edmund Optics) before entering the detection module. For the experiments presented in Fig. 2, a 50× and 0.95-numerical-aperture (NA) (MPlanApo, Olympus) objective was utilized. As high numerical aperture was not a primary concern for the results in Fig. 3, we characterized the light field without microscope. For Fig. 4, this was substituted with a 20× and 0.5-NA (UPlanFL, Olympus) objective to accommodate the clear aperture of the 60-μm-diameter metalens sample. Together, the objective and tube lens form a microscope that images the sample onto a phase-only spatial light modulator (SLM) (X15223-16, Hamamatsu) within the detection module. Following the microscope, the beam polarization is precisely curated through the coordinated action of a half-wave plate (HWP) (VR1-532, LBTEK) and a linear polarizer (LP) (LPVISC050-MP2, Thorlabs), facilitating selective polarization analysis. A 50:50 (R:T) beam splitter (BS) (BS1455-A, LBTEK) directs the wavefront toward the active area of SLM. The modulated field is subsequently reflected by the same BS and projected onto a 1080×1440-px CMOS camera (CS165MU, Thorlabs) for intensity acquisition. During characterization, a sequence of quasi-random phase holograms is loaded on the SLM, and the resulting diffraction patterns are recorded by the CMOS camera to serve as the raw data for iterative field reconstruction.
### Modulation Efficiency of the SLM

The complexity of the SLM pattern and its distance ($d$) to the CMOS sensor must be balanced to achieve optimal modulation efficiency. When $d$ is too small, the CMOS captures insufficient information from the SLM pixels, leading to low modulation efficiency and reconstruction failure. However, this can be mitigated by reducing the SLM pattern pixel size. Conversely, an excessively large $d$ restricts the collection of high-frequency components, thereby degrading spatial resolution. Furthermore, to suppress the inherent crosstalk of the SLM liquid crystals, we employ a pixel-binning strategy—combining multiple SLM pixels into a single "macro-pixel"—to reduce local phase gradients. We performed a series of optimization experiments on the reconstruction performance of the USAF target by varying the SLM pattern pixel size and the distance between the SLM and CMOS (see Fig. S6 discussed in part D of the Supplementary Materials), and we determined that $d = 55$mm, and a 5×5 macro-pixel configuration yield the highest modulation performance.

**Encoding phase patterns design**

The encoding phase patterns for the SLM are designed based on two constraints. First, they should incorporate high phase gradients to accommodate the measurement of high-spatial-frequency components. Second, to mitigate the inherent crosstalk effect of the liquid-crystal SLM, the patterns must exhibit local smoothness. To strike a balance between high-resolution modulation and crosstalk suppression, we first generated a 205×255 random phase pattern following a normal distribution within the range of $(-\pi, \pi]$ with a variance of $0.3\pi$. Subsequently, these patterns were upsampled to 1024×1272 using bicubic interpolation. These resulting grayscale images were addressed to the SLM for measurement encoding, shown in Fig. S7 of the Supplementary Materials.

**Design and fabrication of the metalens**

We designed a focusing metalens based on the geometric phase (Pancharatnam–Berry phase) principle, where the local phase modulation is achieved by spatially varying the orientation angles of the nanopillars(*4*). The metalens, designed for an operating wavelength of 532 nm with a focal length of 3.5 mm. The nanopillars, featuring lateral dimensions of 130 nm × 450 nm, are arranged in a square lattice with a period of 450 nm. In the fabrication process, magnetron sputtering was employed to deposit the $TiO_2$ film, while electron-beam lithography (EBL) was used for high-precision patterning.

**Supplementary Materials**

Supplementary Text

Fig. S1. Polarization reconstruction accuracy as a function of sub-region number.

Fig. S2. Polarization reconstruction based on global phase offset.

Fig. S3. Simulated and measured phase profiles of the x-component.

Fig. S4. Reconstructed polarization map of the light field emergent from a metalens using our method.

Fig. S5. Polarization reconstruction accuracy as a function as phase patterns.

Fig. S6. Reconstruction performance of the USAF target under different SLM pattern pixel sizes and varying distances d between the SLM and CMOS.

Fig S7. Encoding phase patterns loaded on SLM (a total of 60).

Table S1. Comparison with recent CDI-based studies

**Acknowledgements**

**Funding:** This work is funded by Zhejiang Provincial Natural Science Foundation of China (LR25F050002).
**Author contributions:** X.H., X.S. and Xin L. conceived the project. X.S. developed the methodology, performed the simulations, and conducted the experiments. X.S. and Xin L. developed the software. X.S., Z.C., and X.H.

optimized the sample preparation protocols and prepared the samples. X.H. and Y.L. acquired the financial support for the project leading to this publication. X.H. and Xu L. supervised the project. All authors contributed to the manuscript preparation. **Competing interests:** The authors declare no conflicts of interest. **Data availability:** Data underlying the results presented in this article are available from the corresponding authors (X.H.) upon reasonable request. **Code availability:** Customized code used in this study is available at https://github.com/hao-laboratory/PS-CMI. The repository is currently private and will be made publically accessible upon acceptance of the manuscript.

## References

1. N. Yu, F. Capasso, Flat optics with designer metasurfaces. *Nat Mater* **13**, 139–50 (2014).

2. D. Frese, Q. Wei, Y. Wang, L. Huang, T. Zentgraf, Nonreciprocal Asymmetric Polarization Encryption by Layered Plasmonic Metasurfaces. *Nano Lett.* **19**, 3976–3980 (2019).

3. M. Cen, J. Wang, J. Liu, H. He, K. Li, W. Cai, T. Cao, Y. J. Liu, Ultrathin Suspended Chiral Metasurfaces for Enantiodiscrimination. *Advanced Materials* **34**, e2203956 (2022).

4. M. Khorasaninejad, W. T. Chen, R. C. Devlin, J. Oh, A. Y. Zhu, F. Capasso, Metalenses at visible wavelengths: Diffraction-limited focusing and subwavelength resolution imaging. *Science* **352**, 1190–1194 (2016).

5. R. C. Devlin, A. Ambrosio, D. Wintz, S. L. Oscurato, A. Y. Zhu, M. Khorasaninejad, J. Oh, P. Maddalena, F. Capasso, Spin-to-orbital angular momentum conversion in dielectric metasurfaces. *Opt. Express* **25**, 377 (2017).

6. M. Zhao, M. K. Chen, Z.-P. Zhuang, Y. Zhang, A. Chen, Q. Chen, W. Liu, J. Wang, Z.-M. Chen, B. Wang, X. Liu, H. Yin, S. Xiao, L. Shi, J.-W. Dong, J. Zi, D. P. Tsai, Phase characterisation of metalenses. *Light Sci Appl* **10**, 52 (2021).

7. C. Zheng, W. Wang, Y. Ji, Y. Hu, S. Zhang, Q. Hao, High-space–bandwidth product characterization of metalenses by information fusion of multi-angle illumination. *Optica* **12**, 374 (2025).

8. B. Liu, J. Cheng, M. Zhao, J. Yao, X. Liu, S. Chen, L. Shi, D. P. Tsai, Z. Geng, M. K. Chen, Metalenses phase characterization by multi-distance phase retrieval. *Light Sci Appl* **13**, 182 (2024).

9. M. G. Mayani, K. R. Tekseth, D. W. Breiby, J. Klein, M. N. Akram, High-resolution polarization-sensitive Fourier ptychography microscopy using a high numerical aperture dome illuminator. *Opt. Express, OE* **30**, 39891–39903 (2022).

10. J. Kim, S. Song, H. Kim, B. Kim, M. Park, S. J. Oh, D. Kim, B. Cense, Y. Huh, J. Y. Lee, C. Joo, Ptychographic lens-less birefringence microscopy using a mask-modulated polarization image sensor. *Sci Rep* **13**, 19263 (2023).

11. X. Liu, S. Tu, Y. Hu, Y. Peng, Y. Han, C. Kuang, X. Liu, X. Hao, In situ fully vectorial tomography and pupil function retrieval of tightly focused fields. *Nat Commun* **16**, 3478 (2025).

12. M. N. Jacobs, Y. Esashi, N. W. Jenkins, N. J. Brooks, H. C. Kapteyn, M. M. Murnane, M. Tanksalvala, High-resolution, wavefront-sensing, full-field polarimetry of arbitrary beams using phase retrieval. *Opt. Express* **30**, 27967 (2022).

13. Y. Wu, M. K. Sharma, A. Veeraraghavan, WISH: wavefront imaging sensor with high resolution. *Light Sci Appl* **8**, 44 (2019).

14. M. Zhao, M. K. Chen, Z.-P. Zhuang, Y. Zhang, A. Chen, Q. Chen, W. Liu, J. Wang, Z.-M. Chen, B. Wang, X. Liu, H. Yin, S. Xiao, L. Shi, J.-W. Dong, J. Zi, D. P. Tsai, Phase characterisation of metalenses. *Light Sci Appl* **10**, 52 (2021).

15. F. Zhang, B. Chen, G. R. Morrison, J. Vila-Comamala, M. Guizar-Sicairos, I. K. Robinson, Phase retrieval by coherent modulation imaging. *Nat Commun* **7**, 13367 (2016).

16. X. Pan, C. Liu, J. Zhu, Coherent amplitude modulation imaging based on partially saturated diffraction pattern. *Opt. Express* **26**, 21929 (2018).